\newcommand{\average}[1]{\langle{#1}\rangle}
\newcommand{\pardiff}[2]{\frac{\partial{#1}}{\partial{#2}}}
\newcommand{\diff}[2]{\frac{d{#1}}{d{#2}}}

\newcommand{\PNo}{\average{P_{1}}}
\newcommand{\PN}{\average{P_{0}}}
\newcommand{\PNot}{\average{P_{1}(t)}}
\newcommand{\PNt}{\average{P_{0}(t)}}
\newcommand{\oeff}{\omega_{\rm eff}}
\newcommand{\geff}{\gamma_{\rm eff}}
\newcommand{\Teff}{T_{\rm eff}}

\documentclass[aps, superscriptaddress, a4paper,prb,amsmath,amssymb, showpacs, twocolumn, floatfix]{revtex4-1}

\usepackage{graphicx, color, subfigure,amsmath}

\usepackage{hyperref}
\hypersetup{colorlinks={true}}

\begin{document}
\title{Non-linear dynamics of a driven nanomechanical single electron transistor}
\author{P. G. Kirton}
\affiliation{School of Physics and Astronomy, University of Nottingham, Nottingham NG7 2RD, United Kingdom}
\affiliation{SUPA, School of Physics and Astronomy, University of St Andrews, KY16 9SS, United Kingdom}
\author{A. D. Armour}
\affiliation{School of Physics and Astronomy, University of Nottingham, Nottingham NG7 2RD, United Kingdom}
\date{\today}
\begin{abstract}
We analyze the response of a  nanomechanical resonator to an external drive when it is also coupled to a single-electron transistor (SET).  The interaction between the SET electrons and the mechanical resonator depends on the amplitude of the mechanical motion leading to a strongly non-linear response to the drive which is similar to that of a Duffing oscillator. We show that the average dynamics of the resonator is well described by a simple effective model which incorporates damping and frequency renormalization terms which are amplitude dependent.  We also find that for a certain range of parameters the system displays interesting bistable dynamics in which noise arising from charge fluctuations causes the resonator to switch slowly between different dynamical states.
\end{abstract}

\maketitle

\section{Introduction}

Nanoelectromechanical systems (NEMS) in which the transport electrons in a mesoscopic conductor are coupled to a nanomechanical resonator\cite{blencowe:cp} have been studied extensively over the last few years. Prominent examples include resonators coupled to tunnel junctions,\cite{mozyrsky:02,flowers:07,bennett:10b} single-electron transistors (SETs)\cite{knobel:03, mozyrsky:04,armour:04a,naik:06} or quantum dots.\cite{zippilli:09,hussein:10,bennett:10,piovano:11} The electro-mechanical interaction in NEMS is typically rather weak and although there are important examples where non-linear coupling plays a significant role,\cite{gorelik:98,koenig:08,weick:10} in many cases  a linear description is sufficient.

When the  electro-mechanical coupling is weak, the effect of the resonator on the average current flowing through the conductor can provide an extremely sensitive measure of the mechanical displacement.\cite{blencowe:cp} However, the electrons also act back on the nanomechanical resonator and in the weak-coupling limit their effect on the resonator is typically analogous to a thermal bath.\cite{mozyrsky:02,blencowe:cp,clerk:05} Fluctuations in the current give rise to Gaussian fluctuations of the mechanical resonator characterized by an effective temperature, which can be quite different from the thermodynamic temperature of the system's surroundings.\cite{mozyrsky:04,armour:04a,naik:06,zippilli:09}

If the electro-mechanical coupling is increased then this simple picture inevitably breaks down and a more complex dynamics emerges.\cite{doiron:06,clerk:05,usmani:07,rodrigues:07a} The resonator dynamics can become highly non-linear, even if the electro-mechanical coupling itself remains linear, and the fluctuations can become non-Gaussian though exactly what happens depends strongly on the type of charge transport involved. For example, for a resonator coupled to a normal-state SET, the electrons always damp the mechanical motion on average,\cite{armour:04a, armour:04b} but the resonator probability distribution nevertheless gradually changes from a Gaussian to a bimodal form as the coupling is increased.\cite{doiron:06,pistolesi:07} In contrast, if the conductor tends to transfer energy to the mechanical resonator (as is the case for an appropriately tuned superconducting SET), increasing the coupling leads to a  transition in the resonator dynamics which change from fluctuations about a
fixed point to a state of self-sustaining oscillations\cite{clerk:05,usmani:07,rodrigues:07a} when the intrinsic damping
of the mechanical resonator is no longer sufficient to balance the energy transferred by the current flowing through the conductor.

When the mechanical component of a NEMS with weak electro-mechanical coupling is driven to large amplitudes  its influence on the charge dynamics of the conductor is greatly enhanced.\cite{chtchelkatchev:04,labadze:10} The resulting change in the charge transport also necessarily affects the way in which the charges act back on the mechanical system leading to a feedback process which can also generate strongly non-linear mechanical dynamics. Such effects have recently been investigated theoretically and seen experimentally in suspended carbon nanotube systems.\cite{steele:09,lassagne:09,nocera:12, meerwaldt:12} The effective enhancement of electro-mechanical coupling in the presence of driving has also been used in a novel form of force microscopy: When a cantilever which is capacitively coupled to quantum dots in a nearby substrate is strongly driven the resulting back-action on the mechanical dynamics can be used to infer information about the electronic structure of the
dots.\cite{bennett:10}

In this paper we use a simple model system consisting of a nanomechanical resonator linearly coupled to a normal state SET\cite{armour:04a, armour:04b, doiron:06} to explore how  even very weak linear electro-mechanical coupling can give rise to a strongly non-linear response when the resonator is driven close to resonance. In the weak-coupling limit and in the absence of driving, the SET acts on the resonator like a thermal bath with an effective temperature proportional to the bias voltage; it also damps the mechanical motion and renormalizes the frequency of the resonator.\cite{armour:04a} We find that for drives above a certain threshold the mechanical response as a function of frequency becomes strongly non-linear and the mechanical system displays many of the characteristics of the Duffing oscillator:\cite{duffing} frequency pulling, a strongly asymmetric line shape, hysteresis and bistability. Exploiting the fact that the underlying electro-mechanical coupling is very weak, we
describe the effect of the SET on the resonator in terms of a simple model which includes damping and frequency renormalization terms which are both amplitude dependent.\cite{rodrigues:07b,bennett:10} We find that a calculation of the average mechanical response as a function of drive frequency using these two quantities leads to results which are in very good agreement with a full Monte-Carlo simulation of the coupled dynamics.

For a range of drive frequencies and amplitudes the average mechanical response we calculate predicts the coexistence of high and low amplitude states. Monte-Carlo simulations show that in most cases the system spends all its time in just one of the two available states (which one depends on the initial state of the resonator). However, we also find that there exists an interesting regime of bistability where the noise in the system is able to shift the system back and forth between the high and low amplitude states even when the two states are still quite different in terms of amplitudes, phases and the average currents flowing through the SET. In this case the switching between the two states is extremely slow compared to all the other time-scales of the system and could be detected in practice through a characteristic enhancement of the low-frequency current noise.

The structure of this paper is as follows. In Sec.\ \ref{sec:model} we outline our model for the driven SET-resonator system and derive master equations for the full coupled dynamics. Next, in Sec.\ \ref{sec:linear}, we describe the behavior in the regime where the mechanical amplitude remains small and the resonator dynamics can be described using an appropriately modified version of the effective thermal bath description which applies to the undriven system.  Then, in Sec.\ \ref{sec:nonlinear}, we analyze the non-linear dynamics which occur when the strength of the drive is increased. We calculate the effective damping and frequency renormalization of the resonator as a function of the amplitude and use these quantities to calculate the average mechanical response as a function of the drive frequency, comparing the results with those obtained from numerical simulations. In Sec.\ \ref{sec:bistab} we investigate a bistable regime where noise induced switching between two
different states of the mechanical system  occurs at a rate which is much slower than the internal dynamics of both the SET and the resonator.  Finally, in Sec.\ \ref{sec:conc}, we present our conclusions. Brief details on the Monte-Carlo simulations are given in the Appendix.

\section{Model system} \label{sec:model}

\begin{figure}
 \centering
\includegraphics[width=0.7\columnwidth]{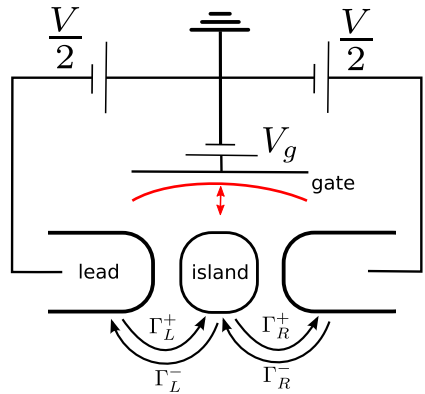}
\caption{(Color online) Circuit diagram of the SET resonator system. The SET is coupled by tunnel junctions to two leads. A gate capacitor is used to tune the operating point, one plate of which is mechanically compliant to provide electro-mechanical coupling.}\label{fig:SET}
 \end{figure}

The SET-resonator system we consider is sketched schematically in Fig.\ \ref{fig:SET}. The SET island is coupled to the left and right leads by tunnel junctions with equal capacitances, $C_J$, and a bias voltage, $V$, is assumed to be applied symmetrically. A gate electrode is used to tune the operating point of the island. The island capacitor consists of one plate which is mechanically compliant, giving rise to a position dependent capacitance, $C_g(x)$. This means that as the plate moves the operating point of the SET changes. Thus, as the charge on the island fluctuates, the electrostatic force on the plate changes, giving rise to electro-mechanical coupling. As long as the displacement of the resonator, $x$, is small compared to the distance, $d$, between the resonator and the SET island when the two are uncoupled,   we can make a linear approximation for the dependence of the gate capacitance on the position of the resonator,\cite{armour:04a}
\begin{equation}
 C_g(x)=C_g^0\left(1-\frac{x}{d}\right),
\end{equation}
where $C_g^0$ is the capacitance when $x=0$.

The dynamics of the SET are determined by the relative sizes of the energy scales in the system: the charging energy of the island, $E_C=e^2/2C_\Sigma$ (where $C_\Sigma$ is the total capacitance of the island\cite{Note1}), the thermal energy, $k_{\rm B}T$, and the energy scale of the bias voltage, $eV$.
The Hamiltonian for the system with $n$-charges on the SET island can be written as\cite{armour:04a}
\begin{equation}
 H_n=E_C\left[n^2-2nn_g^0\left(1-\frac{x}{d}\right)\right]+\frac{p^2}{2m}+\frac{m\omega_0^2x^2}{2}-xF(t),\label{eq:hn}
\end{equation}
where  $n_g^0=C_g^0V_g/e$, $p$ is the resonator momentum, $\omega_0$ and $m$ are the frequency and the mass of the resonator and $F(t)$ is the external drive.  We work in a regime where $E_C\sim eV\gg k_{\rm B} T$, which means that only two charge states are accessible to the system, assuming $0\leq n_g^0\leq1$, these states are $n=0,1$. Note that this simplification does not affect what follows as we do not include the effects of intrinsic non-linearities in the resonator\cite{lifshitz:08} which could in general lead to an explicit dependence of the behavior on the number of excess charges on the island.\cite{Weick:11}

There are four electron tunneling processes which can change the charge state of the island.
We denote the rates for these processes by $\Gamma^\pm_{L(R)}$ where
$+(-)$ represents transitions which go in the same (opposite) direction to the applied bias, and $L(R)$ denotes transitions at the left (right) junction as shown schematically in Fig.\ \ref{fig:SET}.
The rates can be calculated within the {orthodox model},\cite{ferry:09} and in the zero-temperature limit are given by
\begin{equation}
\Gamma^\pm_{L(R)}=\Theta(\Delta E^\pm_{L(R)})\frac{\Delta E^\pm_{L(R)}}{R_Je^2} \label{eq:rates}
\end{equation}
where $\Theta(\cdot)$ is the Heaviside step function and $R_J$ is the junction resistance. The $\Delta E$ terms are the free energy differences associated with each of the transitions,
\begin{gather}
 \Delta E_{L}^\pm=\pm E_{L}\pm m\omega_0^2x_0x, \\
 \Delta E_{R}^\pm=\pm E_{R}\mp m\omega_0^2x_0x,
\end{gather}
 where,
\begin{gather}
 E_L=E_C\left(1-2n_g^0\right)+\frac{eV}{2} \\
 E_R=-E_C(1-2n_g^0)+\frac{eV}{2},
\end{gather}
and we have introduced $x_0=-2n_g^0E_C/m\omega_0^2d$, the change in the equilibrium displacement of the resonator when the SET changes between the charge states $n=0$ and $n=1$.\cite{armour:04a}
The bias voltage term, $eV/2$, accounts for the change in energy of the leads associated with the tunneling of an electron.\cite{ingold:92}

 We now adopt a dimensionless description: We introduce  $\tilde t=t/\tau $, where  $\tau=(\Gamma_L^++\Gamma_R^+)^{-1}=2R_Je^2/eV$, and the scaled position, $\tilde x=x/x_0$.  The dimensionless resonator frequency is $\epsilon=\omega_0\tau$ and the scaled tunnel rates are
\begin{gather}
 \tilde\Gamma_L^\pm=\Theta(\pm\Delta_L\pm\kappa \tilde x)(\pm\Delta_L\pm\kappa \tilde x), \label{eqn:GLpm}\\
 \tilde\Gamma_R^\pm=\Theta(\pm\Delta_R\mp\kappa \tilde x)(\pm\Delta_R\mp\kappa \tilde x), \label{eqn:GRpm},
\end{gather}
where $\kappa=m\omega_0^2x_0^2/eV$ is the dimensionless electro-mechanical coupling strength and $\Delta_{L(R)}=E_{L(R)}/eV$.\cite{armour:04a} 
From now on we use the dimensionless forms of all quantities and so drop the tildes.

The step functions in the expressions for the tunnel rates have important consequences for large $|x|$. If  $x>x_{\rm max}$, where $x_{\rm max}=\Delta_R/\kappa$, then the only possible allowed processes are those in which an electron tunnels onto the island $\Gamma_L^+$ or $\Gamma_R^-$ so at most one tunneling event can occur until $x<x_{max}$. The same kind of effect occurs when $x<x_{\rm min}$, with $x_{\rm min}=-\Delta_L/\kappa$. In this case the only allowed processes are those in which an electron tunnels off the island. Under such circumstances the resonator  blocks transport through the SET, an effect which is essentially the classical counterpart of the Frank-Condon blockade.\cite{doiron:06, pistolesi:07, koch:05}

The above expression for the Hamiltonians and tunnel rates can be used to write down classical master equations which describe the evolution of probability distributions for the state of the SET and the resonator,\cite{armour:04a}
 \begin{gather}
\begin{split} \label{eqn:Lou0}
 \dot P_0(x,v;t)=\{H_0,P_0\}+(\Gamma_R^++\Gamma_L^-)P_{1}-(\Gamma_L^++\Gamma_R^+)P_0,
\end{split}\\
\begin{split}  \label{eqn:Lou1}
 \dot P_{1}(x,v;t)=\{H_{1},P_{1}\}-(\Gamma_R^++\Gamma_L^-)P_{1}+(\Gamma_L^++\Gamma_R^+)P_0,
\end{split}
\end{gather}
where $\{.,.\}$ is a Poisson bracket. Here we have defined the joint probability distributions $P_{0(1)}(x,v;t)$ for the SET to have $n=0(1)$  and the resonator to have position, $x$, and velocity, $v$, at time $t$. The Hamiltonians $H_{0(1)}$ are the dimensionless form of Eq.\ \eqref{eq:hn} with $n=0,1$.

The master equations [Eqs.\ \eqref{eqn:Lou0} and \eqref{eqn:Lou1}] together provide a simple model for the SET-resonator system. The description is readily generalized to include the damping and thermal fluctuations of the mechanical resonator due to its interactions with its surroundings apart from the SET.\cite{armour:04a} However, we shall not include such effects explicitly in our analysis and assume that the interaction with the SET electrons dominates the damping and fluctuations of the mechanical resonator.\cite{naik:06, bennett:10}

\section{Linear response} \label{sec:linear}

When the dynamics of the system never take it into a region where the step functions in the tunnel rates, $\Theta(\cdot)$, are important the system remains linear. This occurs as long as the condition $\kappa|x|\ll\Delta_{L,R}$ is satisfied for all of the phase space explored by the system. In practice these conditions are met provided both the electro-mechanical coupling is weak, $\kappa\ll 1$, and the drive is not too strong.

Within the linear regime the master equations can be approximated as
\begin{multline}
   \dot P_0=[\epsilon^2x-f(t)]\pardiff{P_0}{v}-v\pardiff{P_0}{x}\\+(\Delta_R-\kappa x)P_{1}-(\Delta_L+\kappa x)P_0,
  \end{multline}
  \begin{multline}
 \dot P_{1}=[\epsilon^2(x-1)-f(t)]\pardiff{P_{1}}{v}-v\pardiff{P_{1}}{x}\\-(\Delta_R-\kappa x)P_{1}+(\Delta_L+\kappa x)P_0,
\end{multline}
where $f(t)=F(t)\tau^2/mx_0$.
Hence the full probability distribution of the resonator, $P(x,v;t)=P_0(x,v;t)+P_1(x,v;t)$, evolves according to\cite{armour:04a}
\begin{equation} \label{eqn:PMEnoF}
\dot P(x,v;t)=[\epsilon^2x-f(t)]\pardiff{P}{v}-\epsilon^2\pardiff{P_{1}}{v}-v\pardiff{P}{x}.
\end{equation}
These master equations can be used to find equations of motion for all moments of the position and velocity of the resonator,
\begin{equation}
 \pardiff{}{t}\average{x^nv^m}=\iint dx\,dv x^nv^m\dot P(x,v;t).
\end{equation}

We now analyze the behavior of the system for an external driving force of the form,
\begin{equation}
 f(t)=f_0\sin\omega_D t.
\end{equation}
In the long-time limit the first moments of the mechanical resonator oscillate at the drive frequency, so we make the following ansatz for the fluctuating part of the position,
\begin{equation} \label{eqn:anax}
 \delta x= C{\rm e}^{-i\omega_D t}+C^*{\rm e}^{i\omega_D t},
\end{equation}
where we have defined $\delta x=\average{x}-\average{x}_{ss}$ with $\average{x}_{ss}$ the steady state of the undriven system with $f(t)=0$ (since the force has zero average when integrated over an integer number of periods). The equations of motion  for the first moments of the system can be written as
\begin{gather}
 \delta\dot x=\delta v, \\ \delta \dot v=\epsilon^2(\delta P-\delta x)+f(t), \\ \delta \dot P=\kappa \delta x-\delta P,
\end{gather}
where $\delta v=\average{v}$ and $\delta P=\average{P_1}-\average{P_1}_{ss}$ with $\PNo=\iint dx dv P_{1}$. Substituting the ansatz Eq.\ \eqref{eqn:anax}, into the expression for $\delta \dot P$, and taking the Fourier transform we obtain
\begin{equation}
  \delta P(\omega)=\frac{\kappa}{1-i\omega}\left[C\delta(\omega-\omega_D)+C^*\delta(\omega+\omega_D)\right],  \label{eq:pnfs}
\end{equation}
hence
\begin{equation}
 \delta P(t)=\frac{\kappa}{1+\omega_D^2}(\delta x-\delta v).\label{eq:pnrs}
\end{equation}
This allows us to write,
\begin{equation}
 \delta \ddot x=-\epsilon^2\left(1-\frac{\kappa}{1+\omega_D^2}\right)\delta x-\frac{\epsilon^2\kappa}{1+\omega_D^2}\delta v+f(t), \label{eqn:SHO}
\end{equation}
which is simply the equation of a driven harmonic oscillator with renormalized frequency and damping due to the SET. These are given by,
\begin{align} \label{eqn:goeffF}
 \oeff^2=\epsilon^2\left(1-\frac{\kappa}{1+\omega_D^2}\right), && \geff=\frac{\epsilon^2\kappa}{1+\omega_D^2}.
\end{align}
These quantities take on a very similar form to those in the undriven system,\cite{armour:04a} but now it is the frequency of the drive which enters these expressions in place of the natural frequency of the resonator.\cite{rodrigues:05}

Solving Eq.\ \eqref{eqn:SHO} we obtain the coefficient from our ansatz, Eq.\ \eqref{eqn:anax},
\begin{equation}
 C=\frac{if_0(\oeff^2-\omega_D^2)-f_0\geff\omega_D}{2(\oeff^2-\omega_D^2)^2+2\geff^2\omega_D^2}. \label{eq:cfn}
\end{equation}
The amplitude of the position oscillation of the resonator is then given by,
\begin{equation}\label{eqn:Ampx}
A_x=2|C|.
\end{equation}
To quantify the limits on the linear theory we introduce the critical amplitude,
\begin{equation}
A_c=\frac{x_{\rm max}-x_{\rm min}}{2}=\frac{1}{2\kappa}.
\end{equation}
As long as $A_x<A_c$ the response as a function of frequency is a Lorentzian, centered around the renormalized  frequency of the resonator.\cite{Note2}

 For an undriven resonator the fluctuations in the  position and velocity are Gaussian in the weak coupling limit and can be described by invoking an effective temperature.\cite{armour:04a} For the driven case a similar result is found, but the effective temperature is not a constant. The equations of motion for the variances (e.g.\ $\delta x^2=\average{x^2}-\average{x}^2$) take the form,
\begin{gather}
 \delta\dot {x^2}=2\delta xv, \label{eqn:ODEx2}\\
 \delta\dot {v^2}=2\epsilon^2(\delta v_{1}-\delta xv).
\end{gather}
This set of equations is completed by,
\begin{gather}
 \delta \dot{xv}=\epsilon^2(\delta x_{1}-\delta x^2)+\delta v^2, \\
 \delta \dot {x}_0=\delta v_N-\kappa\delta x^2+\Delta_R \delta x_{1}-\Delta_L\delta x_0, \\
 \delta \dot {x}_{1}=\delta v_{1}+\kappa\delta x^2-\Delta_R \delta x_{1}+\Delta_L\delta x_0, \\
 \delta \dot {v}_{0}=-\epsilon^2\delta x_0-\kappa\delta xv +\Delta_R \delta v_{1}-\Delta_L\delta v_0-\epsilon^2\PNo\PN,  \label{eqn:ODEvN0}\\
\delta \dot {v}_{1}=-\epsilon^2\delta x_{1}+\kappa\delta xv -\Delta_R \delta v_{1}+\Delta_L\delta v_0+\epsilon^2\PNo\PN, \label{eqn:ODEvN1}
\end{gather}
where $\average{P_0}=1-\average{P_1}$ and we have defined, for example, $\delta x_0=\iint dx dv P_0x-\average{x}\PN$. Although these equations can be solved analytically we do not give the solution here as it is rather cumbersome.

The dynamics of the variances is simplest in the adiabatic limit where $\omega_D\ll \geff$. In this case the resonator can be thought of as relaxing to a thermal distribution at each point during the drive cycle. Under these conditions one can find the steady state of Eqs.\ \eqref{eqn:ODEx2}---\eqref{eqn:ODEvN1} assuming in the first instance that $\PNo$ is quasi-static. In this case we find that the fluctuations in the position and velocity of the resonator can be described, in essentially the same way as the undriven case, by an effective thermal distribution with
\begin{equation} \label{eqn:Teff}
 \frac{k_{\rm B}\Teff}{eV}=\PNot\PNt,
\end{equation}
though now the term $\PNot\PNt$ oscillates according to Eqs.\ \eqref{eq:pnfs} and \eqref{eq:cfn}. The oscillations in the SET charge therefore generate oscillating components in the effective temperature (and hence the variances $\delta x^2$ and $\delta v^2$) at $\omega_D$ and $2\omega_D$. When $\omega_D\sim \geff$ the oscillations get smaller as the resonator cannot relax fast enough to follow the oscillations in the effective temperature. If the drive is much faster than $\geff$ then the oscillations are washed out and the variances of the resonator are set by the effective temperature averaged over the drive period.

\begin{figure}
 \centering
{\includegraphics[width=0.48\columnwidth]{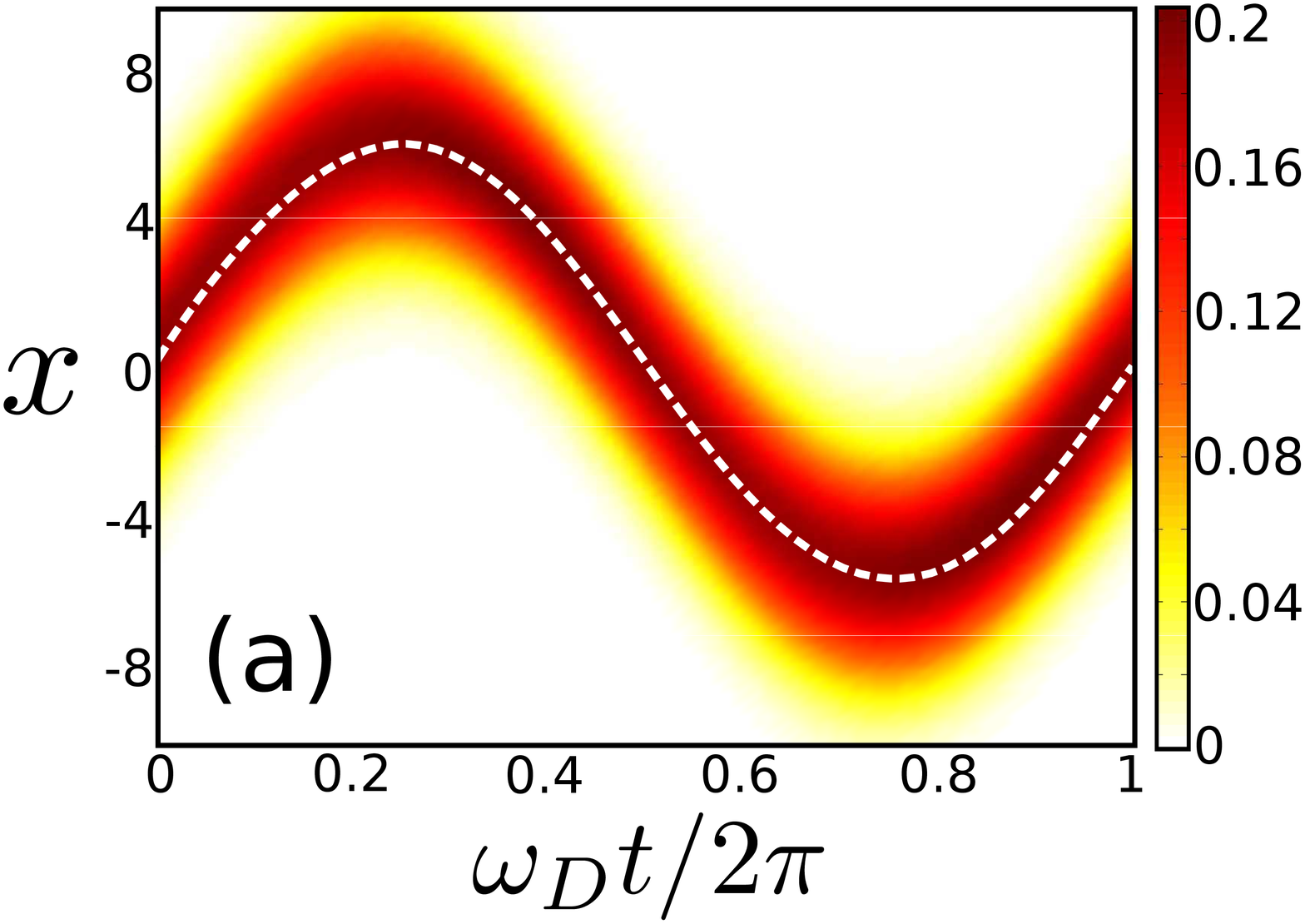} }
 {\includegraphics[width=0.44\columnwidth]{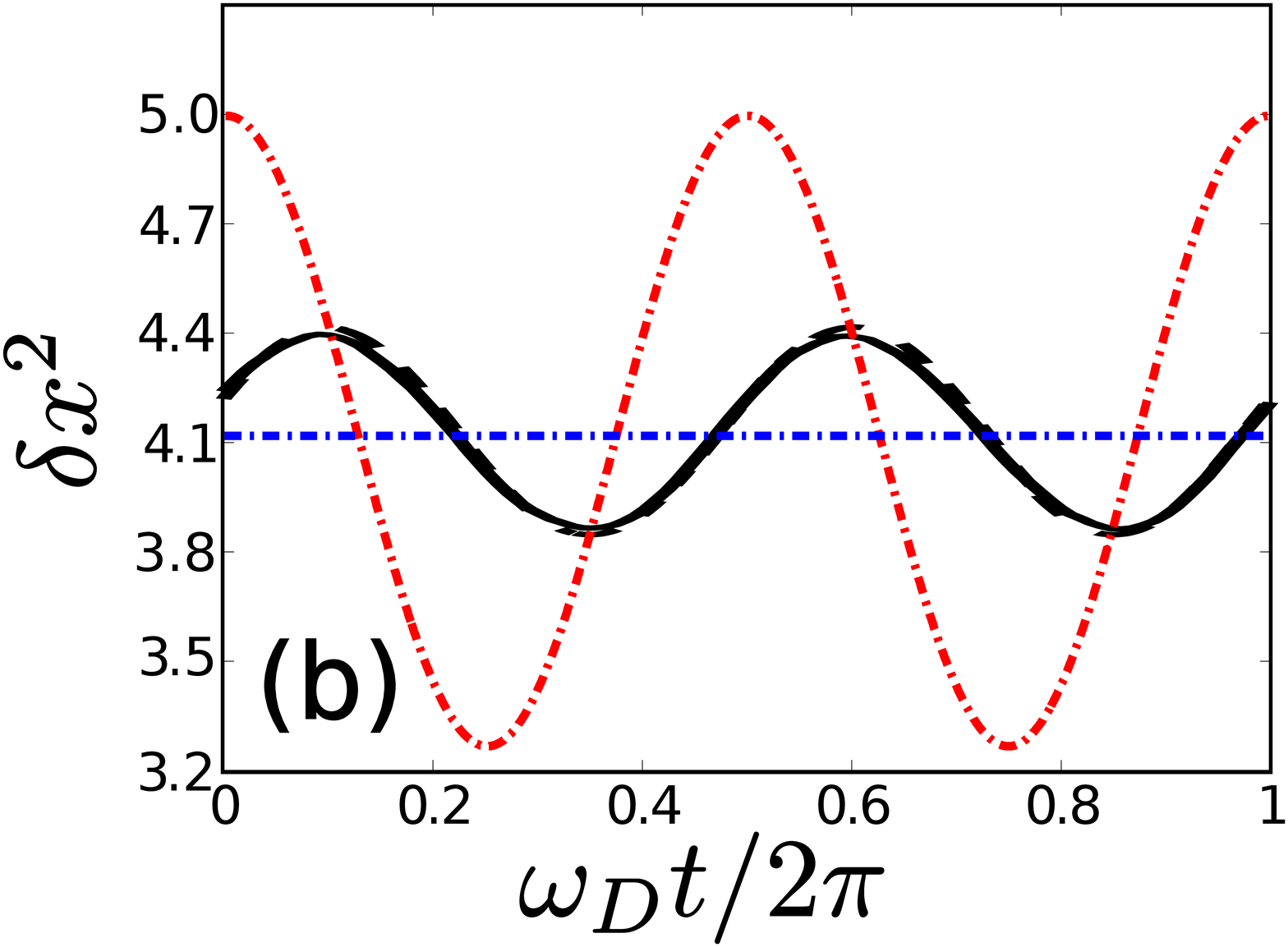}}
\caption{(Color online) Oscillations in both the mean and variance of the position distribution over a period of the drive. The parameters are $\epsilon=0.3$, $\kappa=0.05$, $f_0=0.5$, and $\omega_D=0.002\pi$. In (a) we show the full position distribution (obtained from a Monte-Carlo calculation) and the analytic solution for the mean as the dashed (white) line. In (b) we show the variance; the solid (black) line is the numerical result, and the dashed (black) line shows the solution to Eqs.\ \eqref{eqn:ODEx2}-\eqref{eqn:ODEvN1}. For comparison we also show as the dash-dotted (red) line the variance expected for an adiabatically slow force, from Eq.\ \eqref{eqn:Teff}, and the horizontal line shows the variance corresponding to the average temperature experienced over the cycle. Note that for all numerical calculations we chose $\Delta_L,\Delta_R$ so that the system is at the charge degeneracy point, $\PNo=1/2$, where the oscillations in $\PNo\PN$ at frequency $\omega_D$ vanish.}\label{fig:variances}
\end{figure}

The analytical results in the linear regime can be compared with Monte-Carlo simulations (see the Appendix for more details) of the system dynamics.\cite{doiron:09}
An example of the full position distribution over one period of the drive is shown in Fig.\ \ref{fig:variances}(a). As expected, the mean is well described by the linear analytic result from Eq.\ \eqref{eqn:anax}. In Fig.\ \ref{fig:variances}(b) we compare the numerically simulated variance in the distribution over one drive period to the solution of Eqs.\ \eqref{eqn:ODEx2}---\eqref{eqn:ODEvN1}. The two curves show good agreement, the small deviations occur because a few of the numerical trajectories enter the non-linear region.  The behavior of the variances in the limits of very fast, $\omega_D\gg\geff$, and very slow, $\omega_D\ll\geff$, driving forces are also shown in Fig. \ref{fig:variances}(b) as a horizontal (blue) line and a dot-dashed (red) curve, respectively. For the parameters of the simulation $\omega_D/\geff\sim 1.4$, leading to results which lie between the two limits.

\section{Non-linear response} \label{sec:nonlinear}

Even if the electro-mechanical coupling is extremely weak, $\kappa\ll 1$, a slight increase in the drive strength can be enough to reach a regime in which the amplitude of the oscillations exceeds $A_c$. Thus, the resonator explores regions of phase space where the effects of the Frank-Condon blockade are important. When this happens the tunneling processes in the SET start to become significantly modified by  the step function constraints, the linear theory breaks down and a richer, more complex dynamics emerges. In this section we will examine the response of the resonator to a drive which is strong enough to lead to non-linear behavior and develop techniques to describe the dynamics in this regime.
We start with a very simple approximate way of including the effects of the step functions. This approach provides an intuitive description of the non-linear dynamics of the resonator which is qualitatively correct. A more detailed calculation of the amplitude dependence of the damping and frequency is then presented which is in good quantitative agreement with Monte-Carlo simulations.

\subsection{Reduced coupling approach}

When the system is outside of the linear region, the effective damping and frequency shift of the resonator will be modified and the expressions in  Eq.\ \eqref{eqn:goeffF} are no longer valid. The main effect of the step functions in the tunnel rates is to block electrons from traveling through the SET and when electrons cease to flow their back-action on the resonator will stop too. Assuming that {\it all} electron tunneling stops whenever $x>x_{\rm max}$ or $x<x_{\rm min}$,\cite{doiron:06, pistolesi:07} we can account for the consequent reduction in the damping and frequency shift simply by weighing the effective coupling strength by the proportion of time the resonator spends inside this region.\cite{doiron:09}

We start by assuming that the position of the resonator varies harmonically at the frequency of the drive,\cite{Note3}
\begin{equation} \label{eqn:ansatzxt}
 \delta x(t)=A_x\sin(\omega_D t),
\end{equation}
where we recall $\delta x=\average{x}-\average{x}_{ss}$. Using this position dependence and the assumption that the influence of the SET on the resonator is switched off when $x>x_{\rm max}$ or $x<x_{\rm min}$, we can define an amplitude dependent effective coupling,
\begin{equation}
 \kappa_A=\frac{\kappa\omega_D}{2\pi}\int_0^{\frac{2\pi}{\omega_D}}\Theta\left(\Delta_L+\kappa \average{x(t)}\right)\Theta\left(\Delta_R-\kappa \average{x(t)})\right)\, dt,
\end{equation}
which is readily integrated to give
\begin{equation} \label{eqn:kappaA}
 \kappa_A =
  \begin{cases}
   \frac{2\kappa}{\pi} \arcsin\left(\frac{1}{2\kappa A_x}\right)&  A_x \geq \frac{1}{2\kappa} \\
   \kappa       &  A_x < \frac{1}{2\kappa}
  \end{cases}
\end{equation}
The renormalized coupling gives rise to an implicit amplitude dependence in the damping and frequency shift,
\begin{align} \label{eqn:goeffA}
 \oeff^2(A_x)=\epsilon^2\left(1-\frac{\kappa_A}{1+\omega_D^2}\right), && \geff(A_x)=\frac{\epsilon^2\kappa_A}{1+\omega_D^2}.
\end{align}

\begin{figure}
 \centering
 {\includegraphics[width=0.7\columnwidth]{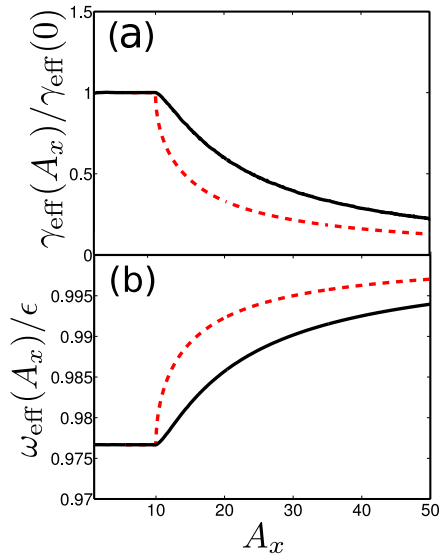} \label{fig:gammaA}}
\caption{(Color online) Amplitude dependent (a) damping and (b) frequency of the resonator. Solid (black) lines show the solution using Eq.\ \eqref{eqn:goeffAint} dashed (red) lines show the results using Eq.\ \eqref{eqn:goeffA}. The parameters used are $\epsilon=0.3$, $\kappa=0.05$, $f_0=0.015$, and $\omega_D=0.29$.}\label{fig:gammaomegaA}
\end{figure}

Examples of these functions are shown as dashed lines in Fig.\ \ref{fig:gammaomegaA}. In the region where the amplitude is still in the linear region, $A_x<A_c$, the damping and frequency shift retain their linear values. Above $A_c$ we see the damping decrease towards zero and the frequency move towards the bare frequency, $\epsilon$. In the large amplitude limit the system spends less and less time inside the linear region where the SET electrons damp the motion. We note that at sufficiently large amplitudes the damping of the resonator due to sources other than the SET electrons (which we neglected) will necessarily  stabilize the dynamics even if, as we assume here, such damping is very small compared to $\geff(0)$.

Solutions for the amplitude of the oscillations are obtained self-consistently from the expression
\begin{equation}\label{eqn:selfconsis}
 A_x=\frac{f_0}{\sqrt{(\oeff^2(A_x)-\omega_D^2)^2+\geff^2(A_x)\omega_D^2}},
\end{equation}
which is derived from the equation of motion for a driven resonator with amplitude dependent damping and frequency.
Equation \eqref{eqn:selfconsis} can have either one or three solutions depending on the choice of parameters.
To find which of these solutions are {stable}\cite{strogatz:01} we derive an equation of motion for $A_x$.\cite{rodrigues:07b} This is given by,
\begin{equation} \label{eqn:AxEoM}
 \dot{A_x}=\frac{1}{A_x}\left[\delta x\dot{\delta x}+\frac{\delta v \dot {\delta v}}{\omega_D^2}\right].
\end{equation}
 The problem is greatly simplified if we assume $A_x$ varies much more slowly than the drive force (to be expected when the coupling is weak).
  We therefore introduce a period averaged amplitude,\cite{rodrigues:07b}
\begin{equation}
 {\tilde{A}_x}=\frac{\omega_D}{2\pi}\int_0^{\frac{2\pi}{\omega_D}} A_x\,dt,
\end{equation}
which obeys the equation of motion,
\begin{equation} \label{eqn:Axdot}
 \diff{\tilde {A}_x}{t}=g(\tilde A_x),
\end{equation}
where,
\begin{equation} \label{eqn:gAx}
 g(A)=-\frac{\geff(A)}{2}\left[A-\frac{f_0^2}{A\left[(\oeff^2(A)-\omega_D^2)+\geff^2(A)\omega_D^2\right]}\right].
\end{equation}
The fixed point solutions are given by $g(A_{fp})=0$ which is just Eq.\ \eqref{eqn:selfconsis}.

The fixed point amplitudes are stable when\cite{strogatz:01, rodrigues:07b}
\begin{equation}
 \left.\diff{g}{A_x}\right|_{A_x=A_{\rm fp}}<0.
\end{equation}
Hence we find that where Eq.\ \eqref{eqn:selfconsis} has three solutions:  the small amplitude solution is in the linear region (and hence stable), the intermediate amplitude solution is unstable and the large amplitude solution is a new stable fixed point for the system (where the damping and frequency shift are reduced from the linear values). The presence of more than one stable amplitude is common in driven non-linear systems, such as the Duffing oscillator.\cite{strogatz:01} However, in contrast to the standard Duffing oscillator\cite{strogatz:01} where the nonlinearity arises from the potential (and the damping is always linear), the interaction with the SET charge leads to both non-linear damping and frequency terms. Whether or not the kind of non-linear damping we see in this system is likely to be important for a wide range of NEMS devices is not yet clear. Whilst non-linear damping has also been shown to be very important in the case of a strongly driven AFM cantilever coupled to a dot,\cite{
bennett:10} a recent analysis of the non-linear response arising in carbon nanotube experiments\cite{steele:09} was able to explain the observed behavior using only conservative nonlinearities.\cite{meerwaldt:12}

\begin{figure}
 \centering
 {\includegraphics[width=0.7\columnwidth]{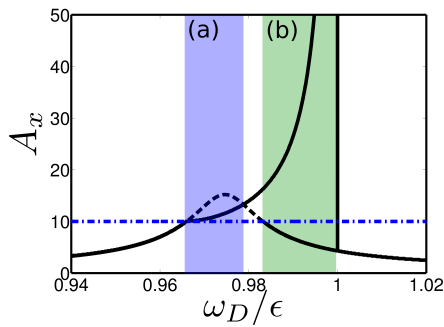}}
\caption{(Color online) The solid (black) lines show the stable solutions to the self-consistency expression, Eq.\ \eqref{eqn:selfconsis}, as a function of drive frequency, for $f_0=0.02$, $\epsilon=0.3$ and $\kappa=0.05$. We also show the linear result as the dashed (black) curve and the amplitude at which the linear theory fails, $A_c$, as the horizontal dot-dashed (blue) line. The shaded regions (a) and (b) are discussed in the text.}\label{fig:stablesimple}
\end{figure}

Using the non-linear damping and frequency shift, along with the conditions on stability of the solutions to Eq.\ \eqref{eqn:selfconsis}, we obtain the response of the system, $A_x$, as a function of drive frequency. An example of the resulting curve can be seen in Fig.\ \ref{fig:stablesimple}. At low frequencies (below the shaded region labeled (a) in Fig.\ \ref{fig:stablesimple}) and at high frequencies $\omega_D>\epsilon$, the system remains in the linear regime; the response is below $A_c$. However, the response to the drive becomes stronger closer to resonance.
In the shaded region labeled (a) in Fig.\ \ref{fig:stablesimple}, the amplitude grows beyond $A_c$, and so the linear and non-linear calculations give different results. The linear calculation leads to a Lorentzian peak centered around $\oeff(0)$. However, in the non-linear case, the frequency shift becomes smaller (leading to a larger effective frequency) for $A_x>A_c$. This means that the drive frequency is farther from resonance than in the linear case, and hence the amplitude is smaller in shaded region (a).


In the shaded region of Fig.\ \ref{fig:stablesimple} labeled (b), the drive frequencies are close to (but below) $\epsilon$ and a high amplitude solution exists because of a positive feedback mechanism. In this regime, as the amplitude grows beyond $A_c$, the enhancement in the effective frequency brings the system closer to resonance with the drive (and the damping decreases as the amplitude grows). However, the system does eventually stabilize  when the amplitude becomes large enough that the system starts to move away from resonance, when $\oeff(A_x)$ starts to increase beyond $\omega_D$.


 For drive frequencies larger than $\epsilon$, the high amplitude solution no longer exists. To see why, consider starting in the high amplitude state and then increasing $\omega_D$ so it is slightly larger than $\epsilon$. If the damping and frequency shift were not amplitude dependent the system would simply oscillate at a slightly lower $A_x$, due to being more detuned from resonance. However, as the amplitude drops, the effective frequency shifts farther from resonance ($\oeff$ decreases with decreasing amplitude) and the effective damping increases, reducing the amplitude of the oscillations farther: The system spirals back down to the linear branch. 

\subsection{Amplitude dependent damping and frequency}
The simple argument presented above is able to give qualitative agreement with the numerical simulations, but to gain quantitative agreement we need to use a more accurate method to calculate the effective damping rate and renormalized frequency.\cite{rodrigues:07b, bennett:10} We assume that a constant amplitude solution to the full non-linear expression exists, and use this  to map the equations onto those of a damped, driven harmonic oscillator. We can then identify the  terms which correspond to an amplitude dependent damping and frequency shift.

We begin with the set of coupled equations for the first moments, including the non-linear position dependence of the tunnel rates (which enters through the step functions), but decoupling the second moments which are present, so that, for example, we assume $\average{x_1}=\average{x}\PNo$.\cite{rodrigues:07b}  We also approximate averages of a function of position, $\average{f(x)}$, by the equivalent function of the average $f(\average{x})$. These approximations have been found to work well for a range of similar systems.\cite{rodrigues:07b, harvey:08, bennett:10}  We therefore obtain
\begin{eqnarray}
 \ddot{\average{x}}&=&\epsilon^2(\average{P_1}-\average{x})+f(t), \label{eqn:xddotnonlin} \\
  \dot{\average{P_1}} &=& \Gamma_L^+(\average{x})(1-\average{P_1})+\Gamma_R^+(\average{x}) \average{P_1}. \label{eqn:Ptnonlin}
\end{eqnarray}
Back-tunneling (i.e., processes described by $\Gamma_{L(R)}^-$) is neglected as its contribution remains very small even within the non-linear regime.\cite{doiron:09}

We now solve for $\PNot$, by integrating Eq.\ \eqref{eqn:Ptnonlin} numerically using the same ansatz for $\average{x(t)}$ as Eq.\ \eqref{eqn:ansatzxt}. Making a comparison between Eq.\ \eqref{eqn:xddotnonlin} and the driven oscillator equation we identify
\begin{equation} \label{eqn:idennonlin}
 -\epsilon^2\average{P_1}=\geff \dot{\average{x}}+\Delta\oeff^2\average{x},
\end{equation}
where we have defined the frequency shift,
\begin{equation}
 \Delta\oeff^2=\oeff^2-\epsilon^2.
\end{equation}
Multiplying Eq.\ \eqref{eqn:idennonlin} by either $\cos\omega_D t$ or $\sin\omega_D t$, and integrating over one period of the drive, we obtain,\cite{rodrigues:07a}
\begin{subequations}\label{eqn:goeffAint}
\begin{gather}
 \geff=-\frac{\epsilon^2}{\pi A}\int_0^{2\pi/\omega_D}  P_1(t)\cos \omega_D t \, dt, \\
 \oeff^2=\epsilon^2\left(1-\frac{\omega_D}{\pi A}\int_0^{2\pi/\omega_D}  P_1(t)\sin \omega_D t \, dt\right).
\end{gather}
\end{subequations}
These expressions correspond to the usual physical interpretation of the damping and frequency shift arising from the SET electrons: The damping is due to the \textit{out of phase} (with respect to $x$) component of the average island charge, while the frequency shift is due to the \textit{in phase} component.\cite{rodrigues:07a, bennett:10,pistolesi:07}

In Fig.\ \ref{fig:gammaomegaA} we plot the amplitude dependence of the damping and frequency shift calculated using Eq.\ \eqref{eqn:goeffAint}, along with the simple results calculated previously using Eq.\ \eqref{eqn:goeffA}. We see that Eq.\ \eqref{eqn:goeffAint} is always larger than the simple estimates. This is because the damping does not simply switch off as soon as the mean resonator position travels through the non-linear boundary as some electron tunneling processes can still occur even  when  $\average{x}<x_{\rm min}$ or $\langle x\rangle >x_{\rm max}$.

\begin{figure}
 \centering
 {\includegraphics[width=0.7\columnwidth]{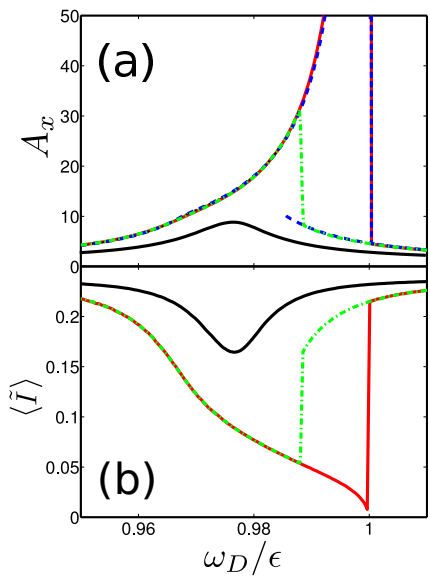}}
\caption{(Color online) Amplitude of the resonator (a) and average current (b) as a function of drive frequency. Results from a Monte-Carlo simulation with a forward (solid (red) line) and backward (dot-dash (green) curve) frequency sweep for $f_0=0.025$ are shown together with  the fully linear results for $f_0=0.01$ (solid (black) curve). Notice that for each sweep the initial conditions for a given frequency were chosen to match the final state found for the previous value of the frequency. The results obtained using amplitude dependent effective damping and frequency shift [Eq.\ \eqref{eqn:goeffAint}] are also shown in (a) plotted as a  dashed (blue) curve. The other parameters used are $\epsilon=0.3$ and $\kappa=0.05$.}\label{fig:compareananum}
\end{figure}

The amplitude dependent damping and frequency shift can be used in the self-consistency expression, Eq.\ \eqref{eqn:selfconsis}, to find the amplitude of the stable solutions. We now compare the predicted stable amplitudes to Monte-Carlo results for parameters which enter the non-linear region. Since the system has two stable solutions with different amplitudes the dynamics are more complicated than in the linear case. Figure \ref{fig:compareananum} compares the amplitudes obtained using the effective damping and frequency shift with results obtained from Monte-Carlo simulations. In general there is very good agreement, though there is a small region where the low amplitude solution predicted from the amplitude dependent damping and frequency is not seen in the numerics. This small discrepancy is a result of the fluctuations in the system which are included in the Monte-Carlo calculation. These fluctuations allow the system to escape from the low amplitude state to the high amplitude state while the low 
amplitude
state is still stable.

The numerical results are obtained by either increasing or decreasing the frequency progressively and in each case
taking the state of the system after a long trajectory at a given frequency as the initial condition for the next frequency value. Calculated in this way the results show clear hysteresis: when the frequency is swept forwards (from lower to higher values) the resonator always follows the high amplitude branch. However, when  frequency is swept downwards (from higher to lower frequency) the system remains in the low amplitude state for a long time because although a high amplitude solution exists it is too far away from the low amplitude one to be reached by fluctuations on the time-scale of the simulations. However, as the drive frequency is progressively reduced the amplitude of the high amplitude oscillations decreases and that of the low amplitude oscillations increases. Eventually the fluctuations in the low amplitude state are enough to take the resonator out of the linear regime and the low amplitude state is no longer seen.

 Figure \ref{fig:compareananum}(b) shows the current (averaged over a large integer number of periods) as a function of drive frequency for both linear and non-linear cases obtained from Monte-Carlo simulations. For the linear case the current is suppressed in the region where the resonator is resonantly driven; as the amplitude of the oscillations increases one of the tunnel rates is suppressed leading to an overall reduction in current (the system is tuned to a gate voltage which for the undriven steady state position corresponds to a current maximum, the charge degeneracy point). The current flowing through the SET reflects the non-linear behavior of the resonator and there is a clear dependence on the direction of sweep. For $\omega_D\simeq\epsilon$ the current is reduced almost to zero in the forward frequency sweep: The amplitude of oscillation of the resonator is so large that only one tunnel event occurs each time the resonators passes through the region where transport is allowed.


\section{Bistability} \label{sec:bistab}

In contrast to the calculations of the effective damping and frequency shift the Monte-Carlo simulations capture the fluctuations in the resonator's dynamics. For the parameters used in Fig.\ \ref{fig:compareananum} fluctuations take the system from a low to a high amplitude state at a particular drive frequency, but the system is never able to switch back and forth between states of different amplitude. However, it is possible to find parameters where bistability does occur by adjusting the drive amplitude and frequency so that the low and high amplitude states are close enough that fluctuations can carry the system between the states in both directions. Bistability is common to a wide range of damped, driven non-linear systems, such as in the Duffing oscillator,\cite{strogatz:01} but is also seen in other non-linear NEMS such as a resonator coupled to a superconducting SET.\cite{rodrigues:07a, rodrigues:07b}

\begin{figure}
 \centering
 {\includegraphics[width=0.7\columnwidth]{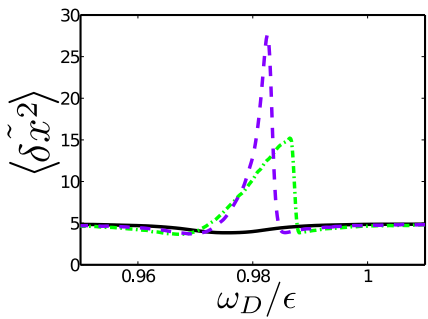}}
\caption{(Color online) Variance of the $x$ distribution  averaged over an integer number of drive periods, $\average{\tilde{\delta x^2}}$, for various drive strengths, the solid (black) curve is $f_0=0.01$ (linear), the dashed (purple) curve is $f_0=0.015$ (bistable) and the dot-dashed (green) curve is $f_0=0.02$ (non-linear). These results were obtained from Monte-Carlo simulations in which the frequency was swept downwards.}\label{fig:bistabvariances}
\end{figure}

\begin{figure}
 \centering
 {\includegraphics[width=0.7\columnwidth]{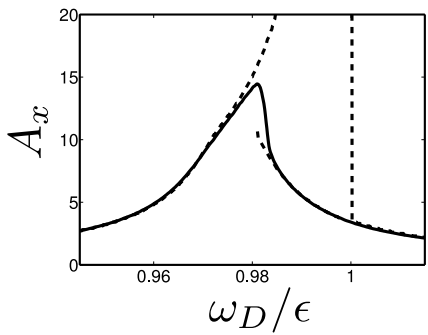}}
\caption{Response of the system as a function of driving frequency; all parameters are the same as in Fig.\ \ref{fig:stablesimple}, except $f_0=0.015$. The solid line shows the numerical response, and the dashed line shows the analytically predicted stable solutions using the damping and frequency shift from Eq.\ \eqref{eqn:goeffAint}. The system shows a bistability in the region where the linear solution reappears.}\label{fig:bistableresponse}
\end{figure}

The fluctuations in the state of the resonator provide important information about its dynamics.  The variance of the position averaged over a large integer number of drive periods, $\langle \tilde{\delta x^2}\rangle$, is shown in Fig.\ \ref{fig:bistabvariances} for different drive strengths. At low drives the dynamics are linear and the variance is approximately constant, with a shallow dip around the resonance.\cite{Note4} The fluctuations are typically much larger for higher drive strengths when the system enters the non-linear regime as can be seen from the dot-dashed curve in Fig.\ \ref{fig:bistabvariances} which  corresponds to the sweep from high to low frequency in Fig.\ \ref{fig:compareananum}. However, by far the largest fluctuations are seen for the
dashed curve in Fig.\ \ref{fig:bistabvariances}
which corresponds to an intermediate drive strength.  A sharp peak in the variance suggests bistability, since a bimodal probability distribution typically gives rise to a very large variance.
Figure \ref{fig:bistableresponse} shows the average amplitude for the same parameters which give rise to the sharp peak in  Fig.\ \ref{fig:bistabvariances}. For this drive strength, in contrast to the behavior seen in Fig.\ \ref{fig:compareananum}(a), there is a smooth crossover, independent of the direction of the frequency sweep, between the two different states predicted by the effective damping and frequency calculation.  This smooth crossover in amplitude at $\omega_D/\epsilon\sim 0.98$ and the corresponding peak in the fluctuations strongly suggest that the system has a bistable region and we can confirm that this is indeed the case by examining individual Monte-Carlo trajectories.

\begin{figure}
 \centering
 {\includegraphics[width=0.7\columnwidth]{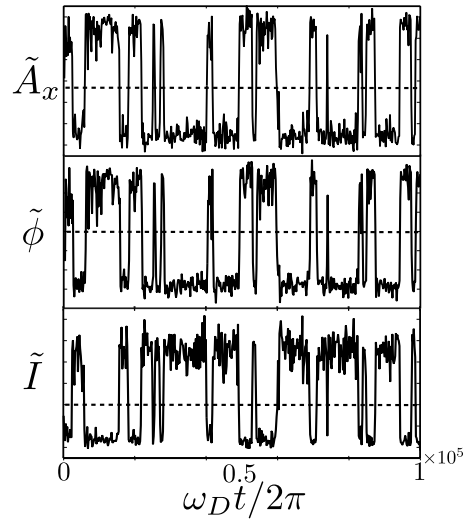}}
\caption{Example trajectories of the amplitude and phase of the resonator along with the current through the SET as a function of time. The dashed lines show  the point at which we choose to split the data into the high and low amplitude states. The parameters are $\epsilon=0.3$, $\kappa=0.05$ and $f_0=0.016$. The drive frequency is chosen such that the occupation probabilities of the two states are equal. Each point in the plots is averaged over 200 periods of the drive.}\label{fig:currentphasetrace}
\end{figure}

Examples of the dynamics during a single trajectory are shown in Fig.\ \ref{fig:currentphasetrace}, where we show a trace of the resonator's amplitude and phase as well as the current through the SET. To make the behavior clear each point in the trajectories is averaged over a timescale which is long compared to the drive period, but short compared to the rate at which the system moves between the two different states. It is clear that the dynamics are well described by a two state model, the system switches between two distinct values for amplitude, phase and current as a function of time. This allows us to split the trajectory into regions which correspond to the high and low amplitude oscillations of the resonator. The phase and current trace are
well correlated with the amplitude of the oscillations: When the system is in the high amplitude state the current is small and the phase is large; the opposite is true for the low amplitude state.

\begin{figure}
 \centering
 {\includegraphics[width=0.9\columnwidth]{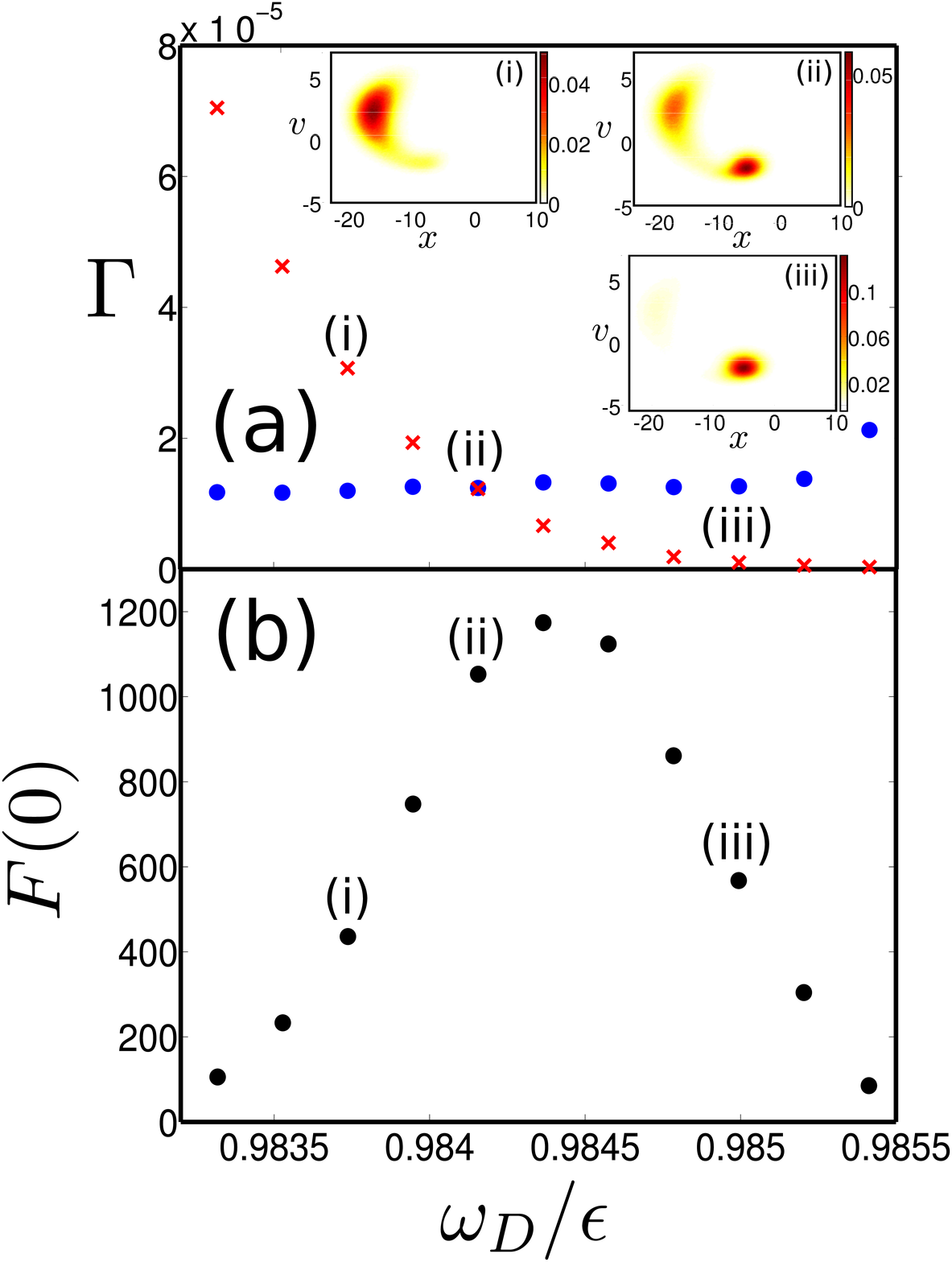}}
\caption{(Color online) (a) Transition rates between the two states as a function of drive frequency, close to the bistability. The (blue) dots denote $\Gamma_{hl}$ the rate from the high amplitude state to the low amplitude state, and the (red) crosses show $\Gamma_{lh}$. The insets show the resonator probability distributions at the drive frequencies labeled (i), (ii) and (iii) in the main panel. The parameters at point (ii) are the same as in Fig.\ \ref{fig:currentphasetrace}. (b) Corresponding zero-frequency Fano factor, calculated from the two state model, as described in the text. The points labeled (i), (ii), and (iii) match those in (a).}\label{fig:ratesfull}
\end{figure}

The transition rates between the two bistable states, $\Gamma_{hl}$ and $\Gamma_{lh}$, where $h(l)$ labels the high (low) amplitude state, are given by
\begin{align}
 \Gamma_{hl}=\frac{1}{\tau_h} && \Gamma_{lh}=\frac{1}{\tau_l}
\end{align}
where $\tau_{h(l)}$ is the average amount of time spent in the state $h(l)$. These rates\cite{Note6} are readily extracted from the Monte-Carlo trajectories. Figure \ref{fig:ratesfull}(a) shows the behavior of $\Gamma_{hl(lh)}$  as the drive frequency is swept through the bistability.  The rates are much slower than all other timescales in the problem, for example  $\geff\sim 0.005$  for the low amplitude state and $\geff\sim 0.002$ for the high amplitude state. This means that the system has time to relax in each of the metastable states and a two state model should be a good approximation to the dynamics on intermediate time-scales.\cite{flindt:05b, usmani:07} The insets to Fig.\ \ref{fig:ratesfull} show probability distributions for the resonator at the three marked drive
frequencies. These are obtained stroboscopically by taking points
at a particular drive phase.  We see that when $\Gamma_{hl}<\Gamma_{lh}$, illustrated in inset (i), the majority of the distribution is in the high amplitude state, which has a wide distribution in phase. At frequencies above the bistability [inset (iii)] the distribution is dominated by the low amplitude state (which is essentially the linear solution here), since $\Gamma_{hl}>\Gamma_{lh}$. Between these two limits [inset  (ii)] the two rates are approximately equal, $\Gamma_{hl}\approx\Gamma_{lh}$, and the distribution contains significant contributions from both states.

The noise in the current flowing through NEMS is known to provide important information about the dynamics of the system.\cite{flindt:05b, usmani:07, harvey:08, bruggemann:12} Assuming that the current is measured on a timescale slower than the drive, the relevant quantity is the period averaged current\cite{camalet:04}
\begin{equation}
 \tilde{I}(t)=\frac{\omega_D}{2\pi}\int_t^{t+\frac{2\pi}{\omega_D}}{I(t')}\,dt'.
\end{equation}
It is the current defined in this way which shows the bistable behavior seen in Fig.\ \ref{fig:currentphasetrace}. Close to zero frequency  the noise in this current will be dominated by the switching rate between the two states, since these are the slowest timescales in the dynamics.\cite{flindt:05b, usmani:07, harvey:08} The noise can the  be quantified by the zero-frequency Fano factor of the current,
\begin{equation}
 F(0)=\frac{\int_{-\infty}^{\infty}\average{\tilde I(t)\tilde I(0)}\,dt}{\average{\tilde{I}}}.
\end{equation}
Assuming the system is well-described by a two-state model it is possible to find a simple analytic expression for $F(0)$:\cite{flindt:05b, usmani:07, harvey:08}
\begin{equation} \label{eqn:fano}
 F(0)=\frac{2P_hP_l\left(\average{\tilde{I}_h}-\average{\tilde{I}_l}\right)^2}{\Gamma_{hl}\average{\tilde{I}_h}+\Gamma_{lh}\average{\tilde{I}_l}},
\end{equation}
where $P_{h(l)}$ and $\average{\tilde{I}_{h(l)}}$ are the occupation probability and average current of the high (low) amplitude state respectively. These quantities can be calculated in a straightforward manner from traces like those in Fig.\ \ref{fig:currentphasetrace}. The predictions for the behavior of the Fano factor based on the two-state description are shown in Fig.\ \ref{fig:ratesfull}(b). As is typical for this kind of dynamics,\cite{flindt:05b, usmani:07, harvey:08, bruggemann:12} we find a large enhancement in the noise close to the bistability: switching between the two metastable states causes large fluctuations in the current. The maximum in the Fano factor occurs for a drive frequency slightly larger than that for which the occupation probabilities of the two states are equal, (the point (ii) in Fig.\ \ref{fig:ratesfull} where $\Gamma_{hl}=\Gamma_{lh}$). As the frequency is increased the difference between the current in the two states, $|\average{\tilde{I}_h}-\average{\tilde{I}_l}|$
increases, as can be seen in Fig.\ \ref{fig:compareananum}(b); this contribution shifts the maximum in $F(0)$ slightly away from the point where $P_h=P_l$.

\section{Conclusions} \label{sec:conc}

We have used a simple model system consisting of a mechanical resonator linearly coupled to a normal state SET to explore the non-linear dynamics which can arise in driven NEMS. Despite weak (linear) electro-mechanical coupling the resonator's dynamics become non-linear when it is driven sufficiently strongly. At large enough amplitudes the effect of the resonator on the SET charge dynamics can no longer be accounted for by a linear correction to the tunnel rates and the charge transport is strongly modified. The modified charge dynamics leads to changes in the damping and frequency shift induced by the SET on the resonator leading in general to an amplitude dependence of these quantities. Such an amplitude dependence is generic in non-linear oscillators and leads to the familiar phenomena of asymmetric frequency response, hysteresis and bistability.

We have focused on a simple model in which the effects of finite temperature on the electron tunneling and the effects of the resonator's surroundings beyond the SET electrons are ignored. Whilst the model can be adapted to include all of these and other complicating factors it is nevertheless useful to work with a simplified description as it lays bare the mechanisms which give rise to non-linear behavior.

For weak, linear, electro-mechanical coupling in NEMS one generally expects the electrical transport to act on the resonator like an effective thermal bath, giving rise to damping of the mechanical motion and Gaussian fluctuations.\cite{mozyrsky:02,armour:04a,naik:06,clerk:05}
The cases where coupling between the electrons and the resonator instead gives rise to negative damping have become established as well-known exceptions to this paradigm.\cite{clerk:05,rodrigues:07a,usmani:07} The way in which the effective thermal description breaks down as the electro-mechanical coupling is increased has also been investigated carefully.\cite{doiron:06,pistolesi:07} Our work has explored a third case: in general one expects driving of the resonator in a NEMS device to lead (via the interaction with the transport electrons) to non-linear dynamics which also fall well outside the effective thermal bath description.\cite{meerwaldt:12}

\begin{acknowledgments}
We thank F. Pistolesi for helpful comments and acknowledge funding from EPSRC (UK) under Grant No.\ EP/I017828/1.
\end{acknowledgments}

\appendix

\section*{Appendix: Monte-Carlo simulations}

In this appendix we briefly summarize the  numerical techniques used to obtain the Monte-Carlo results discussed in the main body of the article. The basic idea is to simulate the dynamics obtained from Eq.\ \eqref{eq:hn} via Hamilton's equations,
\begin{eqnarray}
\dot{x}&=&v \label{eq:aa}\\
\dot{v}&=&-\epsilon^2(x-n)+f(t) \label{eq:ab}.
\end{eqnarray}
 When electron tunneling is taken into account these equations are essentially stochastic, because the island charge, $n$, fluctuates between zero and unity according to the tunnel rates given by Eqs.\ \eqref{eqn:GLpm} and \eqref{eqn:GRpm}.

 To perform Monte Carlo simulations of the trajectory of the system governed by these equations we evolve time in discrete steps of length $\Delta t$. At each time step, a uniform random number, $r\in[0,1]$, is generated and compared with the appropriate rates to see whether the charge state should be updated. For example, if $n=0$ and $r$ is in the interval $[0,(\Gamma_L^++\Gamma_R^-) \Delta t]$ then the state of the SET island is changed $n=1$, otherwise no change is made. The  resonator is then evolved over the timestep using Eqs.\ \ref{eq:aa} and \ref{eq:ab}.


The results shown in the main body of the article are obtained by appropriate averaging over single long trajectories. Starting from  an arbitrary set of initial conditions the trajectory is evolved a time, $t_0$, long enough that the system has lost all memory of the initial condition (this is tested by ensuring that the results are not sensitive to changes in $t_0$). We then evolve the system along the trajectory recording running averages as required.

\end{document}